\documentstyle[emulateapj]{article}

\slugcomment{To appear in PASP (2002 Dec)}

\begin{document}

\title{The Progenitor of Supernova 1993J Revisited\footnote{Based on
observations made with the NASA/ESA {\sl Hubble Space Telescope}, obtained in
part from the data archive of the Space Telescope Science Institute, which is
operated by the Association of Universities for Research in Astronomy, Inc.,
under NASA contract NAS 5-26555.}}

\author{Schuyler D.~Van Dyk}
\affil{IPAC/Caltech, Mailcode 100-22, Pasadena CA  91125}
\affil{email: vandyk@ipac.caltech.edu}
\authoremail{vandyk@ipac.caltech.edu}

\author{Peter M.~Garnavich}
\affil{Physics Department, University of Notre Dame, Notre Dame, IN 46556-5670}
\affil{email: pgarnavi@miranda.phys.nd.edu}
\authoremail{pgarnavi@miranda.phys.nd.edu}

\author{Alexei V. Filippenko}
\affil{Department of Astronomy, University of California, Berkeley, CA 
94720-3411}
\affil{email: alex@astro.berkeley.edu}
\authoremail{alex@astro.berkeley.edu}

\author{Peter H\"oflich}
\affil{Astronomy Department, University of Texas, Austin, TX 78712}
\affil{e-mail: pah@astro.as.utexas.edu}
\authoremail{pah@astro.as.utexas.edu}

\affil{and}

\author{Robert P. Kirshner, Robert L. Kurucz, and Peter Challis}
\affil{Harvard-Smithsonian Center for Astrophysics, Cambridge, MA 02138}
\affil{email: (rkirshner, rkurucz, pchallis)@cfa.harvard.edu}
\authoremail{(rkirshner, rkurucz, pchallis)@cfa.harvard.edu}

\begin{abstract}

From {\sl Hubble Space Telescope\/} images with $0{\farcs}05$ resolution we
identify four stars brighter than $V = 25$ mag within $2{\farcs}5$ of SN 1993J
in M81 which contaminated previous ground-based brightness estimates for the
supernova progenitor.  Correcting for the contamination, we find that the
energy distribution of the progenitor is consistent with that of an early
K-type supergiant star with $M_V \approx -7.0 \pm 0.4$ mag and an initial mass
of 13--22 $M_\odot$.  The brightnesses of the nearby stars are sufficient to
account for the excess blue light seen from the ground in pre-explosion
observations.  Therefore, the SN 1993J progenitor did not necessarily have a
blue companion, although by 2001, fainter blue stars are seen in close
proximity to the supernova.  These observations do not strongly limit the mass
of a hypothetical companion.  A blue dwarf star with a mass up to 30 $M_\odot$
could have been orbiting the progenitor without being detected in the
ground-based images.  Explosion models and observations show that the SN 1993J
progenitor had a helium-rich envelope.  To test whether the helium abundance
could influence the energy distribution of the progenitor, we calculated model
supergiant atmospheres with a range of plausible helium abundances.  The models
show that the pre-supernova colors are not strongly affected by the helium
abundance longward of 4000~\AA, and abundances ranging between solar and 90\%\
helium (by number) are all consistent with the observations.

\end{abstract}

\keywords{supernovae: general---supernovae: individual (SN 1993J)---galaxies:
individual (M81, NGC 3031)---stars: evolution}

\section{Introduction}

Direct identification of the progenitor star has been scarce for the more than
2000 known historical supernovae (SNe); hence, our knowledge of the types of
stars that become SNe is based primarily on models of stellar evolution and
observations of the environments in which SNe occur.  Pre-SN stars have been
detected for only five events: SN~1987A in the LMC (e.g., Walborn et al.~1987),
SN~1961V in NGC 1058 (Goodrich et al. 1989; Filippenko et al. 1995), SN~1978K
in NGC 1313 (Ryder et al.~1993), SN~1997bs in NGC 3627 (Van Dyk et al.~1999),
and SN~1993J in M81 (Filippenko 1993a; Aldering, Humphreys, \& Richmond 1994,
hereafter AHR; Cohen, Darling, \& Porter 1995).  Interestingly, the properties
of all five of these SNe are atypical of type II supernovae (SNe~II).  In fact,
the highly unusual properties of SN 1961V and SN 1997bs (Van Dyk et al.~2000)
suggest that they were actually eruptions of extremely massive stars, similar
to $\eta$~Carinae, rather than genuine SNe.

SN~1993J was initially classified as a SN~II, because its optical spectra
showed hydrogen lines (Filippenko 1993b; Garnavich \& Ann 1993), but it soon
developed the characteristics of a SN~Ib (Filippenko \& Matheson 1993;
Filippenko, Matheson, \& Ho 1993; Filippenko, Matheson, \& Barth 1994) ---
hence, the classification as Type IIb.  Several groups (see the review by
Wheeler \& Filippenko 1996) concluded that the progenitor of SN~1993J was a
massive star (10--20~M$_\odot$) that had lost most of its hydrogen envelope
before exploding.  The models generally employ a close companion star to strip
the progenitor of its envelope, since mass loss through a wind is not efficient
at such low initial masses. The companion gains mass and may reach more than
15~M$_\odot$ before the supernova explosion (Woosley et al. 1994).  H\"oflich,
Langer, \& Duschinger (1993) offered an alternative model, in which a single,
very massive star loses its envelope through a wind before exploding; the
results, however, do not fit the observations as well as do binary star models.

The photometric properties of the SN 1993J progenitor were estimated by AHR,
using an extensive ground-based photographic and CCD archive for M81.  Their
analysis was affected by the range in quality of the observations and the
crowded field near the SN.  AHR concluded that the progenitor was a K0
supergiant with contamination from one or more blue stars projected near the
SN and possibly light from an OB association that spawned SN 1993J.

From a set of CCD images obtained in the 1980s to search for novae in M81,
Cohen et al. (1995) determined that the SN 1993J progenitor was not variable at
the $0.2$ mag level. They also confirmed the AHR estimate of the progenitor
color.  Crotts (1995) obtained high-quality CCD images 600 days after maximum
brightness and found a red star with $I \approx 22.7$ mag within 1\arcsec\ of
SN~1993J.  This star was responsible for the elliptical appearance of the
progenitor in the AHR set of images, even under good seeing conditions.

In the course of the Supernova Intensive Study (SINS) collaboration,
multiwavelength {\sl Hubble Space Telescope\/} ({\sl HST}) images had been
obtained of SN~1993J and the surrounding star field in 1994 and 1995, while the
SN was fading.  The SN and the field have also been observed in 2001, though
not as the main target, by program GO-9073 (see Liu, Bregman, \& Seitzer 2002).
In this paper, we use all these data to help resolve some questions that were
raised by the ground-based data about the progenitor.  In particular, we
estimate the contamination of the progenitor light by stars which are
unresolved in the ground-based images and attempt to determine the source of
the excess blue light seen by AHR.  We also estimate the brightness, color,
age, and initial mass of the progenitor.  In \S 2 we describe the observations
and photometric reduction.  The implications of the {\sl HST\/} data for the
pre-SN observations are discussed in \S 3, and conclusions are presented in \S
4.

\section{Observations and Data Reduction}

Short exposures were obtained, when the SN was still fairly bright, on 1994
April 18 (UT is used throughout this paper) with the {\sl HST\/} WFPC2 through
filters F336W, F439W, F555W, F675W, and F814W, which approximate standard
Johnson-Cousins $UBVRI$ bandpasses.  After the SN faded further, images were
obtained on 1995 January 31 through the same filters, and additionally, the UV
filter F255W. Exposures through filters F450W, F555W, and F814W were also
obtained by GO-9073 on 2001 June 4.  Table 1 summarizes all of the {\sl HST}
observations.

Photometry of these images was performed using HSTphot version 1.1 (Dolphin
2000a,b).  This package automatically accounts for WFPC2 point-spread function
(PSF) variations and charge-transfer effects across the chips, zero-points,
aperture corrections, etc., and returns $UBVRI$ magnitudes as output.
HSTphot appears to have a limiting (image-to-image) photometric
accuracy of $\sim$0.02 mag (Dolphin 2000a).  The transformations, even
at faint magnitudes, should be at this level or better (Dolphin 2000b).

Approximately 25 stellar sources are within $2{\farcs}5$ of the SN that can be
identified in the sum of all images obtained in 1995 (see Figure 1).  Of these,
only six have $B < 25$ mag, and only four have $V < 25$ mag.  For
identification purposes, the angular distances and position angles from
SN~1993J for stars A--F have been listed in Table~2.  The magnitudes from each
epoch, along with the adopted magnitudes for the stars in the SN environment
(the uncertainty-weighted average of the three epochs), are given in Table 2.
Measurements for the brightness of SN~1993J itself are given in Table~3.

A distance to M81 of 3.63 $\pm$0.34 Mpc ($\mu = 27.80 \pm 0.20$ mag) has been
determined from {\sl HST\/} observations of Cepheid variables (Freedman et
al.~1994). The reddening to SN~1993J is not well-determined. Richmond et
al.~(1994) estimate $E(B-V) = 0.08$--0.32 mag. From early-time spectra,
Clocchiatti et al.~(1995) find a visual extinction $A_V = 0.74 \pm 0.05$ mag.
As AHR point out, the range in likely $A_V$ to the SN is about 0--1.5 mag.  We
adopt $A_V = 0.75$ mag, in agreement with Clocchiatti et al. (1995) and
consistent with the midrange of colors and magnitudes of stars in the SN's
environment (see below).  Therefore, a star with $B=25$ mag, if physically
associated with M81, would have absolute magnitude $M_B \approx -3.6$.

The SN was still bright in 1995 January ($V \approx 18.8$ mag) and faint stars
closer than ${\sim}0{\farcs}25$ (${\sim}4.4$ pc at the distance of M81) would
not be resolved, if present.  Stars with $B > 25$ mag, even in 2001, are more
than 11 times fainter in $B$ than the estimated brightness of the progenitor
(see below) and could not substantially affect the progenitor photometry.

The position of the red star reported by Crotts (1995; his estimates are
$\delta\theta=0{\farcs}84$, P.A. = 344\arcdeg) is consistent with Star A in the
{\sl HST\/} images.  However, Crotts' measurement for the star, $I \approx
22.7$ mag, is $\sim 2$ mag fainter than our {\sl HST\/} measurement of 20.7
mag.  Also, the color for this star, $R-I = 1.14$ mag, based on the {\sl HST\/}
measurements, is much redder than the $R-I \approx 0.5$ mag reported by Crotts.
Star A may be somewhat variable, as seen from the magnitudes derived from all
three epochs of {\it HST} observations. On the other hand, the variability is
certainly not of this degree, nor can we easily explain the difference due to
bandpass differences.  This discrepancy demonstrates the difficulty in
obtaining accurate photometry of closely blended stars, even from the best
ground-based sites.

\section{Discussion}

Analysis of archival ground-based images by AHR showed that no single star
could fit the progenitor's energy distribution.  They suggested that the
spectrum is best fit by a slightly reddened K0 supergiant, plus light from an
OB association surrounding the SN.  From the {\sl HST\/} images, we find four
stars, A--D, with $V \lesssim 24$ mag, that fall within the resolution element
of observations typical for AHR (FWHM $\approx 1{\farcs}5$, corresponding to
FWZI ${\approx} ~2{\farcs}5$), which may have influenced flux measurements of
the progenitor by AHR.  Star A appears to be variable in all bands, which might
possibly have influenced the progenitor flux measurements by Cohen et
al.~(1995).  However, they rule out variability for the progenitor at the
$\sim$0.2 mag level; the variability we measure in $V$ for Star A is comparable
to, or slightly larger than, this level, although the measured behavior here is
consistent with that shown by Cohen et al.  (We note that the apparent
variability for Star A we measure in $B$ and $R$ is $>0.2$ mag.)  Two blue
stars are seen in Figure 1 very close to the SN, but they are near the limit of
the detection threshold and are therefore likely too faint to substantially
affect the progenitor photometry. The $V = 23.5$ mag star just east of due
north and the $V = 24.3$ mag, somewhat extended, source to the southwest, both
near the edge of the circle in the figure, are well in the wings of the typical
ground-based seeing disk.  All other stars within the circle are quite faint
($V \approx 25$--26 mag).

Based on a visual inspection of the {\sl HST\/} images, SN 1993J does not
appear to be a member of a cluster or association.  Using IRAF\footnote{IRAF
(Image Reduction and Analysis Facility) is distributed by the National Optical
Astronomy Observatories, which are operated by the Association of Universities
for Research in Astronomy, Inc., under cooperative agreement with the National
Science Foundation.}/DAOFIND, a detailed comparison between the density of
stars within 2\arcsec\ of the SN with the overall star density on the $B$-band
frames also does not indicate a statistically significant stellar clustering
around the supernova.  SN 1993J is at a large radial distance from the center
of M81 ($4{\farcm}8$), located along an outer spiral arm, which is interspersed
with dust lanes.  Nonetheless, the photometry indicates that recent star
formation has occurred in the SN environment.

Figure 2 shows the color-color diagrams for the environment.  The $(U-B)\
vs.\ (B-V)$ diagram serves as the best diagnostic of the reddening and
extinction in the SN's neighborhood, which, within the uncertainties in the
stellar photometry, is $A_V = 0$--1.5 mag, with the midvalue $A_V = 0.75$ mag
[$E(B-V) = 0.24$ mag, assuming a Cardelli, Clayton, \& Mathis (1989) reddening
law].  This is consistent with the locus of most of the stars on this diagram.
We adopt this midvalue, although the reddening in the environment appears to be
variable and may at some locations exceed this value, up to $A_V \approx 2$
mag.  Figure 3 shows the color-magnitude diagrams for the environment.  The
brightest blue star on the diagrams is, of course, SN 1993J, which clearly has
unusual colors, due to its emission-line spectrum.  From these diagrams, Stars
C and D are consistent with late O-type to early B-type, likely on the main
sequence (other possible fainter, young main-sequence stars are also seen in
the environment).  Star A is quite red and is consistent with a K-type
supergiant, while, from Figures 3a and 3b, Star B is either more evolved than
Stars C and D or is experiencing greater reddening.  The detected fainter
stars, with $V > 24.5$ mag, tend to be red, owing to the relative lack of
sensitivity of the $U$ and $B$ bands (see Table 1).

The spectral energy distributions for Stars A--D are compared with that
observed for the SN 1993J progenitor in Figure 4.  The archival images used by
AHR were taken under a wide range of conditions, but, even for those obtained
under the best conditions, the magnitude estimate for the progenitor still
contained some or all of the light from these four stars.  Star A contributed
$\sim$25\% of the $I$-band flux measured by AHR for the progenitor, but a
negligible amount of blue light.  The three other stars, B--D, contributed very
little red light, but together contributed nearly 80\% of the blue light and
may be the source of the apparent excess flux at short wavelengths.  One
concern might be that background subtraction performed by AHR as part of the
measurement process for the ground-based images already accounts for the mean
contribution of these stars in an aperture, since the spatial distribution of
stars within the aperture is similar to the distribution of stars in the
overall background (due to the lack of a well-defined OB association around the
progenitor).  However, presuming that the mode is used to estimate the aperture
photometry background in a sky annulus outside of the aperture, the fainter
pixels in an image are given greater weight than the brighter ones concentrated
in resolved stars.  (Besides, AHR preferred PSF-fitting over aperture
photometry in their measurements.) Therefore, the nearby stars would still
contaminate the PSF of the progenitor star measured from the ground.

To determine the likely energy distribution of the SN~1993J progenitor we must
correct the AHR distribution for this contamination, limits for which are shown
in Figure 5.  The sum of the light from the four stars is computed with (1)
equal weight (dotted line) and (2) following a possibly more realistic case, in
which the contaminating star contribution declines with angular distance from
the progenitor (dot-dashed line).  Because the central part of a ground-based
PSF is similar to a Gaussian, the degree of contamination is likely to decline
with distance in a similar manner.  To weight the contribution from Stars A--D
we use a Gaussian distribution, which is the sum of the unit-weight central
distribution of the progenitor light and the distribution of the light from the
contaminating star,
$$ F_\lambda \; =\; \sum_{i={\rm A...D}} F_{\lambda i}\;
e^{-r^{2}_i/4\sigma^{2}} ,$$
where $r_i$ is the distance of the contaminating star from the progenitor and
$\sigma$ is the width of the distribution [abbreviated ``$\Sigma$(F*g)'' in
Figure 5].  Since, for a Gaussian, the FWHM = 2.35$\sigma$, and the FWHM is a
function of wavelength, $\sigma$ is also wavelength-dependent.  From the AHR
observational data, excluding the $V$ and $R$ measurements of Perelmuter (1993;
AHR note that these measurements are several $\sigma$ above the mean in these
bands), the values for $\sigma$, weighted by the standard deviations in the
measurements, are $\sigma_U = 0{\farcs}85$, $\sigma_B = 0{\farcs}61$, $\sigma_V
= 0{\farcs}69$, $\sigma_R = 0{\farcs}77$, and $\sigma_I = 0{\farcs}53$.  The
corrected progenitor energy distributions are shown in Figure 5 as the
long-dashed line (equal weight) and the short-dashed line (Gaussian weight).

The actual energy distribution of the progenitor star probably falls somewhere
within these limits.  In the ultraviolet (UV) the total flux from the
surrounding stars is likely ${\sim}1\sigma$ from the AHR $U$-band flux.  Thus,
within the errors, the blue stars near the SN could account for all the UV flux
of the progenitor measured by AHR, and the corrected $U$-band flux from the
progenitor is essentially undetermined.  To possibly better estimate the
progenitor's energy distribution at redder bands, we simulated the light of the
progenitor, plus the contamination from the nearby stars, as would be measured
from the ground.

We did this by first subtracting the SN PSF from the 2001 {\sl HST\/} F450W
($B$), F555W ($V$), and F814W ($I$) images (when the SN was the faintest),
employing a TinyTim model PSF (Krist 1995) in each band and the routines
ALLSTAR and SUBSTAR (Stetson 1987, 1992) within IRAF/DAOPHOT.  We then added at
the SN position a fake star, representing the progenitor, using ADDSTAR within
IRAF/DAOPHOT.  We did this in a series of fake star trials, with a range of
different input apparent magnitudes.  These trial images were then convolved
with a two-dimensional Gaussian with the wavelength-dependent values of
$\sigma$ above, to simulate ground-based observing conditions.  (These
convolved images also then contained the light of the contaminating stars.)
The model PSF was also convolved with the same Gaussian and subsequently
applied to each of the convolved trial images in ALLSTAR, to obtain magnitudes
for each of the fake star trials in each band.  We then determined which
convolved fake progenitor magnitude best matched that measured by AHR in each
band for the actual progenitor.  The input magnitude which resulted in this
best match likely provided the best estimate of the actual progenitor's
brightness in each band.

We show in Figures 6a--6c the input fake, or hypothetical, progenitor images,
before convolution, which resulted in the best matches to the AHR magnitudes
after convolution.  [Note the presence of the light echo (Liu et al.~2002;
Sugerman \& Crotts 2002), particularly in the F450W and F555W images.]  The
resulting magnitude estimates for a hypothetical progenitor star are $B \approx
22.4$, $V \approx 21.6$, and $I \approx 19.7$ mag.  In the F450W image the
faint, very nearby Stars E and F are only $\sim 1/4$ the hypothetical
progenitor star's brightness (some negative pixels resulting from
oversubtraction of the SN are also seen).  The fake progenitor in the $I$ band
has nearly the same brightness as the SN itself in the 2001 F814W image and is
only $\sim$1 mag brighter than Star A.  In a sense, these final trial images
simulate pre-SN WFPC2 images, had it been possible for them to have been
obtained.

We show in Figure 7a the final Gaussian-convolved fake progenitor image for
F814W ($I$) as an example (compare this with Figure 6c).  In general, the final
test images in all three bands closely resemble the ground-based images shown
by AHR (their Figure 1).  Note the elliptical shape of the fake progenitor in
the convolved $I$-band image, due primarily to the presence of Star A.  (For
the $I$-band trials, Star A dominated the resulting convolved stellar profile
for a faint fake progenitor, but the profile resolved into two stars as the
trials were increased in brightness.)  We show in Figure 7b the output image
from ALLSTAR, with the fake progenitor subtracted.  Star A becomes far more
obvious after subtraction, although this star, as well as Stars B--D (which are
much fainter at $I$), are all within the seeing disk of the convolved fake
progenitor.  (Compare Figure 7b with Figure 1 in AHR; their PSF-subtracted
$I$-band image, with the progenitor removed, clearly shows the presence of an
object which we suspect is Star A.)  Our simulation demonstrates that the
ground-based measurements, even under good seeing conditions, contained not
only the light of the progenitor star, but also that of the nearby
contaminating stars.

The formal uncertainties in the magnitude estimates for the progenitor, mostly
due to aperture effects, small PSF mismatches, slight PSF oversubtraction,
small disagreements between HSTphot and DAOPHOT results, and transformations to
standard magnitudes, are estimated to be $\sim$0.30 mag in each band.  We have
estimated that variable seeing in $B$ and $V$ (corresponding to the range in
seeing for the data in AHR) leads to an additional uncertainty of $\sim$0.1
mag.  Furthermore, the uncertainties in the $V$ and, particularly, $B$
magnitudes could be somewhat larger, since the faint stars E and F (and
possibly others) very close to the progenitor are comparable to the
progenitor's brightness. Moreover, the relatively blue light echo (Liu et
al. 2002; Sugerman \& Crotts 2002), although faint and almost 2\arcsec\ from
the SN, may contaminate the convolved fake progenitor's brightness in $B$ and
$V$ to some extent.  The magnitude estimates are shown in Figures 3 and 5.  To
within the uncertainties, the simulated magnitudes agree fairly well with the
limits shown in Figure 5, although the consistency is somewhat better in $B$
and $I$ than in $V$.

Also shown in Figure 3 are solar-metallicity isochrones from Bertelli et
al.~(1994) for 8 Myr and 16 Myr, reddened by $E(B-V) = 0.24$ mag and adjusted
for the M81 distance.  The 16 Myr isochrone fits reasonably well the locus of
Star A. If we assume that the SN progenitor and Star A were coeval, this age
provides at least a lower limit to the progenitor's mass, as 16 Myr is roughly
the main-sequence lifetime of an 12--14 $M_\odot$ star (Schaller et al.~1992;
Bertelli et al.~1994).  With this assumption, we would expect that the SN
progenitor was at least as massive as Star A, since Star A is likely a red
supergiant and therefore also will likely explode as a supernova.  The locus of
the fake progenitor star on the diagrams is reasonably well fit, particularly
in the red, by the 8 Myr isochrone, which is roughly the main-sequence lifetime
of a 21--24 $M_\odot$ star.  However, in the blue, the fake progenitor is
somewhat better fit by the blue loop of the 16 Myr isochrones (again, the
actual uncertainties in $B$ and $V$ for the fake progenitor are likely larger
than the formal uncertainties).  Bluer colors for the progenitor would agree
with the evidence that the progenitor lost most of its hydrogen envelope
(Filippenko et al.~1994) and with a possible transition from red to blue
supergiant during the late stages of the progenitor evolution (e.g., Immler,
Aschenbach, \& Wang 2001).

Additionally, we can assign an approximate spectral type and luminosity class
to the uncontaminated energy distribution for the progenitor, through
construction of model atmospheres for supergiants using the ATLAS12 code
(Kurucz 1993).  One can see that the model colors for an early K-type (K0)
supergiant agree reasonably well with the overall distribution.  Because the
progenitor lost most of its hydrogen envelope (e.g., Filippenko et al.~1994)
and the progenitor atmosphere was helium-rich (e.g., Garnavich \& Ann 1994), we
have calculated the K0~I broad-band colors for three helium abundances: solar,
75\% helium, and 90\% helium (by number). As expected, the colors for the
solar-abundance model are very close to those given by Bessell (1990) from the
Vilnius spectra.  Unfortunately, the three models primarily differ in the UV,
where the corrected energy distribution of the progenitor is relatively
undetermined. At longer wavelengths the colors of all three K0~I star models
are in good agreement with the probable energy distribution of the progenitor.

Finally, after accounting for the contamination by nearby stars, we can place a
limit on the brightness of any blue companion to the SN~1993J progenitor.  The
models show that the K0~I progenitor had $U \approx 24$ mag, thus contributing
very little to the flux measured by AHR in that band.  A close hypothetical
progenitor companion with $U < 22.4$ mag, added to the light from Stars A--D,
would cause the total UV flux to exceed the AHR measurement for the progenitor
by ${\sim}1\sigma$.  Conservatively, any progenitor companion could not have
been brighter than $U = 22.0$ mag and still be consistent with those
measurements.  The AHR uncertainty in the $B$ band is small, so adding an
additional star of $B < 23.0$ mag to the ones accounted for here would exceed
the observed flux by ${\sim}2\sigma$ for a K0~I progenitor.  Thus, the
progenitor companion must have been less luminous than $M_B \approx -5.6$ mag,
assuming $A_V = 0.75$ mag. Since O6~V and later-type dwarfs have $M_B > -5.4$
mag ($M_V > -5.0$ mag; Humphreys \& McElroy 1984) and masses of up to 30
$M_\odot$, we find that any hypothetical companion actually could have been of
very high mass and still remained undetected.

\section{Conclusions}

We have used {\sl HST\/} images of SN~1993J after outburst to search for stars
within a ground-based resolution element, which may have contaminated the
archival ground-based observations of the SN progenitor.  We find four stars
within $2{\farcs}5$ of the progenitor which were at least partially included
in the progenitor energy distribution determined by AHR.  Correcting for this
contamination we find the following.

\begin{description}

\item[1)] The progenitor may have had apparent $V = 21.6 \pm 0.3$, $B-V = 0.8
\pm 0.6$, and $V-I = 1.9 \pm 0.5$ mag, implying an absolute magnitude $M_V
\approx -7.0 \pm 0.4$ for $A_V = 0.75$ mag (the uncertainty in the absolute
magnitude includes the estimated photometric uncertainty, plus the uncertainty
in the distance modulus).  The brightness and colors are consistent with those
of a reddened early K-type supergiant, although possibly somewhat too blue in
the blue bands.  The mass of the progenitor can be constrained by the possible
mass for the nearby red supergiant, Star A, i.e., $\sim$13 $M_\odot$, and the
age and mass of a hypothetical progenitor suggested by its brightness and
colors, i.e., $\sim$22 $M_\odot$.  This range, 13--22 $M_\odot$, is consistent
with other estimates for the progenitor mass in the literature (e.g.,
Podsiadlowski et al.~1993; Woosley et al.~1994; Young, Baron, \& Branch 1995;
Iwamoto et al.~1997).

\item[2)] Model atmospheres with a wide range of helium abundances fit the
observed progenitor energy distribution redward of 4000 \AA.  Although at
shorter wavelengths the models show a greater spread in color with helium
fraction, the data are insufficient to differentiate between them.

\item[3)] The blue excess seen by AHR can be fully accounted for by the nearby
stars.  The progenitor was not a member of a significant OB association.  Any
companion to the progenitor must have had $M_B > -5.6$ mag for reasonable
assumptions of the reddening and distance to the SN.  The mass of the companion
is not very strongly constrained, but must be less than 30 $M_\odot$. This is
consistent with the binary mass-transfer scenario, even for conservative mass
exchange (e.g., Podsiadlowski et al. 1993).

\end{description}

SN 1993J appears to be interacting with circumstellar material ejected by the
progenitor during its evolution (e.g., Filippenko et al.~1994; Van Dyk et
al.~1994; Patat, Chugai, \& Mazzali 1995; Matheson et al.~2000).  The energy
from this interaction continues to power the optical light curve, which in 1995
slowed its decline to less than 2 millimag d$^{-1}$ (Garnavich et al.~1995).
Comparing the magnitudes derived from the {\it HST\/} WFPC2 images in 1995 and
2001, the SN is still declining at only ${\sim}$0.2 mag yr$^{-1}$.  Recent ACS
multi-band images have been obtained with {\sl HST\/} by program GO-9353, but
even with the superior sensitivity and resolution of ACS, the SN is likely
still too bright to successfully isolate a companion or other stars in the
immediate SN environment (${\le}0{\farcs}1$--$0{\farcs}2$).  Only when the SN
sufficiently fades, probably not for a few more {\sl HST\/} cycles, can the
high-resolution imaging be used to search for a companion to the SN progenitor
and better determine the progenitor's nature.

\acknowledgements

This work was supported in part by NASA through grants GO-2563, GO-8648,
GO-9114, and AR-8754 from the Space Telescope Science Institute, which is
operated by the Association of Universities for Research in Astronomy, Inc.,
under NASA contract NAS~5-26555.  We thank Greg Aldering for comments that
helped improve this paper.

%\clearpage

\begin{deluxetable}{lcccc}
\tablenum{1}
\tablewidth{17cm}
\tablecaption{Summary of {\sl HST\/} Observations}
\tablehead{
\colhead{Date} & 
\colhead{Band} &
\colhead{Exp.~Time} &
\colhead{$m_{\rm limit}$\tablenotemark{a}} &
\colhead{Program}  \nl
\colhead{(UT)} &
\colhead{} &
\colhead{(s)} &
\colhead{(mag)}  &
\colhead{}}
\startdata
1994 Apr 18 & F336W & 1200 & 23.7 & GO-5480 \nl
            & F439W & 600  & 24.5 &         \nl
            & F555W & 300  & 25.5 &         \nl
            & F675W & 300  & 25.0 &         \nl
            & F814W & 300  & 24.5 &         \nl
1995 Jan 31 & F255W & 2400 & 22.1 & GO-6139 \nl
            & F336W & 1160 & 23.8 &         \nl
            & F439W & 1200 & 25.4 &         \nl
            & F555W & 900  & 26.6 &         \nl
            & F675W & 900  & 26.1 &         \nl
            & F814W & 900  & 25.5 &         \nl
2001 Jun 4  & F450W & 2000 & 26.7 & GO-9073 \nl
            & F555W & 2000 & 26.8 &         \nl
            & F814W & 2000 & 25.2 &         \nl
\enddata
\tablenotetext{a}{These are the approximate HSTphot 3$\sigma$ detection 
limits, in magnitudes, for the PC chip (programs GO-5480 and GO-6139), 
and for the WF4 chip (program GO-9073).  Magnitudes are given in the 
{\sl HST\/} flight system.}
\end{deluxetable}

\clearpage

\begin{deluxetable}{cccccccccc}
\tablenum{2}
\tablewidth{17cm}
\tablecaption{Measurements of Bright Stars Within 2.5\arcsec\ of SN~1993J}
\tablehead{
\colhead{Star} & 
\colhead{Offset} &
\colhead{P.A.\tablenotemark{a}} &
\colhead{Year} &
\colhead{$U$} &
\colhead{$B$} &
\colhead{$V$} &
\colhead{$R$} &
\colhead{$I$} \nl
\colhead{} &
\colhead{$\arcsec$} &
\colhead{$\arcdeg$} &
\colhead{} &
\colhead{(F366W)} &
\colhead{(F439W)\tablenotemark{b}} &
\colhead{(F555W)} &
\colhead{(F675W)} &
\colhead{(F814W)} }
\startdata
A & 0.73 & 348 & 1994 & 23.82(14) & 23.84(27) & 22.63(04) & 21.66(03) & 
20.67(03) \nl
  &      &     & 1995 & \nodata   & 24.38(18) & 22.86(03) & 22.03(02) & 
20.80(01) \nl
  &      &     & 2001 & \nodata   & 23.80(03) & 22.54(01) & \nodata   & 
20.60(01) \nl
  &      &     & AVE. & 23.82(14)\rlap{:} & 23.82(04) & 22.58(01) & 21.84(03) & 
20.70(01) \nl
\nl
B & 1.44 & 184 & 1994 & 23.01(14) & 23.12(11) & 22.92(05) & 23.05(07) & 
23.16(14) \nl
  &      &     & 1995 & 22.94(14) & 23.27(08) & 23.02(03) & 23.12(04) & 
23.06(05) \nl
  &      &     & 2001 & \nodata   & 23.17(02) & 23.12(02) & \nodata   & 
22.84(10) \nl
  &      &     & AVE. & 22.98(14) & 23.17(03) & 23.07(03) & 23.10(05) & 
23.03(07) \nl
\nl
C & 1.22 & 353 & 1994 & 23.03(11) & 23.98(22) & 23.74(08) & 23.67(12) & 
23.69(24) \nl
  &      &     & 1995 & 22.97(12) & 23.76(12) & 23.78(04) & 23.88(08) & 
23.55(09) \nl
  &      &     & 2001 & \nodata   & 23.86(03) & 23.77(03) & \nodata   & 
23.46(11) \nl
  &      &     & AVE. & 23.00(12) & 23.86(04) & 23.77(04) & 23.82(09) & 
23.53(11) \nl
\nl
D & 1.19 & 149 & 1994 & 22.90(10) & 23.78(19) & 23.80(08) & 23.89(14) & 
24.12(29) \nl
  &      &     & 1995 & 23.04(12) & 23.78(11) & 23.96(05) & 24.04(08) & 
23.90(10) \nl
  &      &     & 2001 & \nodata   & 23.78(03) & 23.90(03) & \nodata   & 
24.19(14) \nl
  &      &     & AVE. & 22.96(11) & 23.78(04) & 23.90(04) & 24.00(09) & 
24.01(13) \nl
E & 0.32 & 335 & 1994 & \nodata   & \nodata & \nodata & \nodata & \nodata \nl
  &      &     & 1995 & \nodata   & 23.84(14) & 24.04(10) & 23.68(12) & 
23.82(14) \nl
  &      &     & 2001 & \nodata   & 24.59(23) & 24.32(07) & \nodata & 
\nodata \nl
F & 0.29 & 282 & 1994 & \nodata   & \nodata & \nodata & \nodata & \nodata \nl
  &      &     & 1995 & 24.35(34) & 24.03(13) & 24.87(09) & 23.77(07) &
23.96(11) \nl
  &      &     & 2001 & \nodata   & \nodata & \nodata & \nodata & 
\nodata \nl
\enddata
\tablenotetext{}{Note: The values in parentheses are the $1\sigma$
uncertainties in the last two digits of the magnitudes.}
\tablenotetext{a}{Position angle, degrees from north through east.}
\tablenotetext{b}{The 2001 observations by GO-9073 were made through the F450W, 
rather 
than the F439W, filter.}
\end{deluxetable}

%\clearpage

\begin{deluxetable}{cccccccc}
\tablenum{3}
\tablewidth{17cm}
\tablecaption{{\sl HST\/} Photometry of SN~1993J}
\tablehead{
\colhead{Date} &
\colhead{UV} &
\colhead{$U$} &
\colhead{$B$} &
\colhead{$V$} &
\colhead{$R$} &
\colhead{$I$} \nl
\colhead{(UT)} &
\colhead{(F255W)} &
\colhead{(F366W)} &
\colhead{(F439W)\tablenotemark{a}} &
\colhead{(F555W)} &
\colhead{(F675W)} &
\colhead{(F814W)} }
\startdata
1994 Apr 18  & \nodata & 17.98   & 18.22 & 17.78 & 16.99   & 17.21 \nl
1995 Jan 31  & 17.30   & 18.55   & 19.10 & 18.75 & 17.90   & 18.63 \nl
2001 Jun 04  & \nodata & \nodata & 20.48 & 20.32 & \nodata & 19.75 \nl
\enddata

\tablenotetext{}{Note: The 1$\sigma$ uncertainties in the magnitudes are all
${\le}0.01$ mag.}
\tablenotetext{a}{The 2001 observations by GO-9073 were made through the F450W,
rather than the F439W, filter.}
\end{deluxetable}

\clearpage

\begin{figure}
\figurenum{1}
%\plotone{fg1.ps}
\caption{The sum of the {\sl HST\/} images obtained in 1995 through the F336W,
F439W, F555W, F675W, and F814W filters, showing the environment around SN~1993J
in M81. The circle centered on the supernova has a radius of $2{\farcs}5$,
which is approximately the full-width near zero intensity (FWZI) of a Gaussian
seeing disk with full-width at half maximum (FWHM) = $1{\farcs}5$ (typical
seeing for observations analyzed by AHR).
The six brightest stars in the SN environment are labelled.
The four brightest, A--D, with $V < 24$ mag, most likely contaminated
the ground-based measurements of the SN progenitor by AHR.}
\end{figure}

%\clearpage

\begin{figure}
\figurenum{2}
\plotone{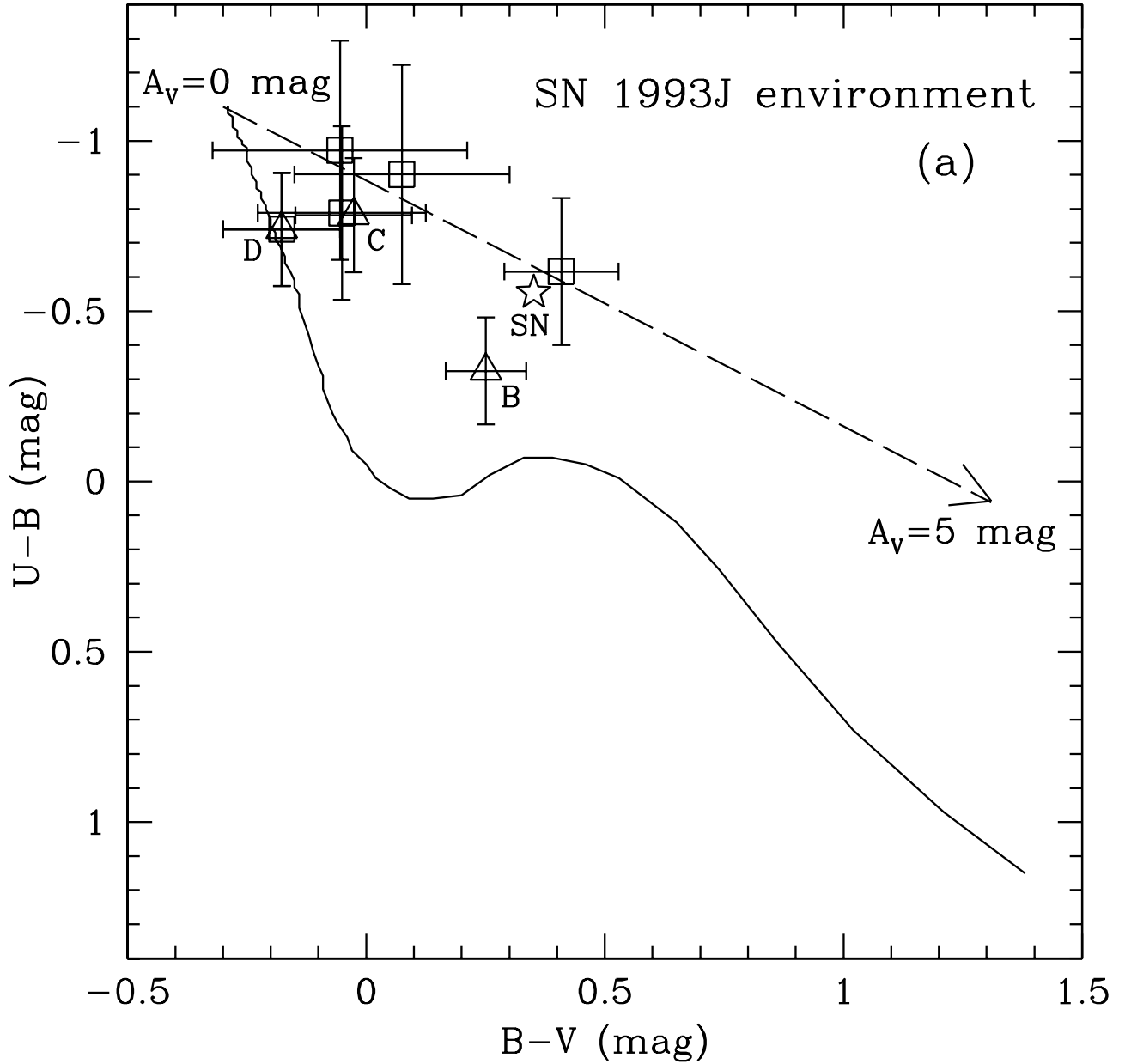}
\caption{The color-color diagrams for the SN 1993J environment (an area
$\sim 2{\farcs}5$ in radius, centered on the SN), measured from
the 1995 {\sl HST\/} images.  {\it (a)} The ($U-B$, $B-V$) diagram; and
{\it (b)} the ($V-R$, $V-I$) diagram.  In both diagrams, Stars A--D are
represented by {\it open triangles} and labelled, whereas {\it open squares\/} 
represent other stars in the environment.  The SN ({\it five-pointed star\/}) 
is also labelled.  In {\it (a)}, Star A is not shown, due to the lack of a 
$U$-band
measurement.  In both diagrams the main-sequence locus is shown ({\it solid
line}), as well as the reddening vector ({\it dashed line}), assuming a
Cardelli et al.~(1989) reddening law.}
\end{figure}

\clearpage

\begin{figure}
\figurenum{2}
\plotone{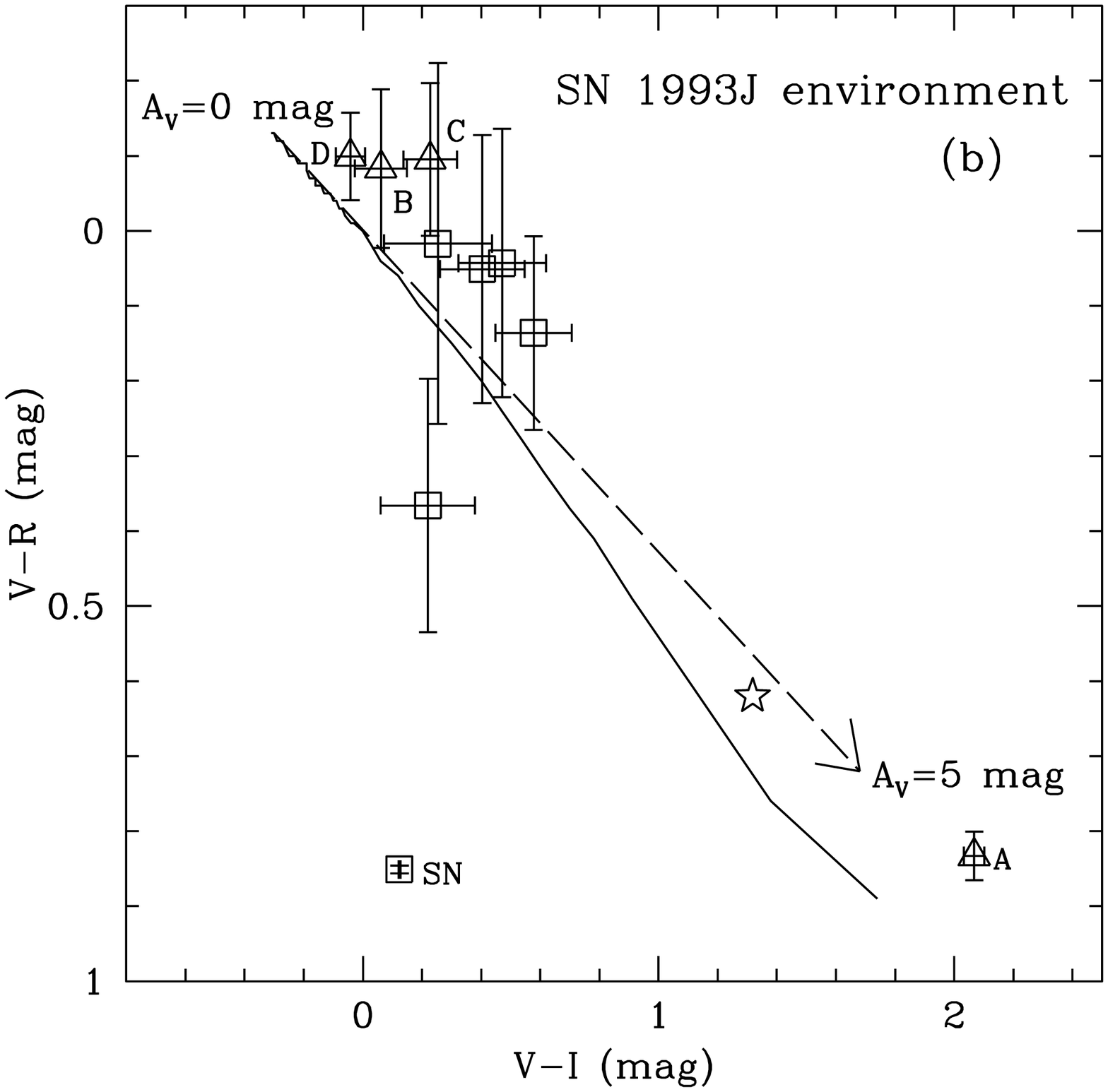}
\caption{(Continued.)}
\end{figure}

\clearpage

\begin{figure}
\figurenum{3}
\plotone{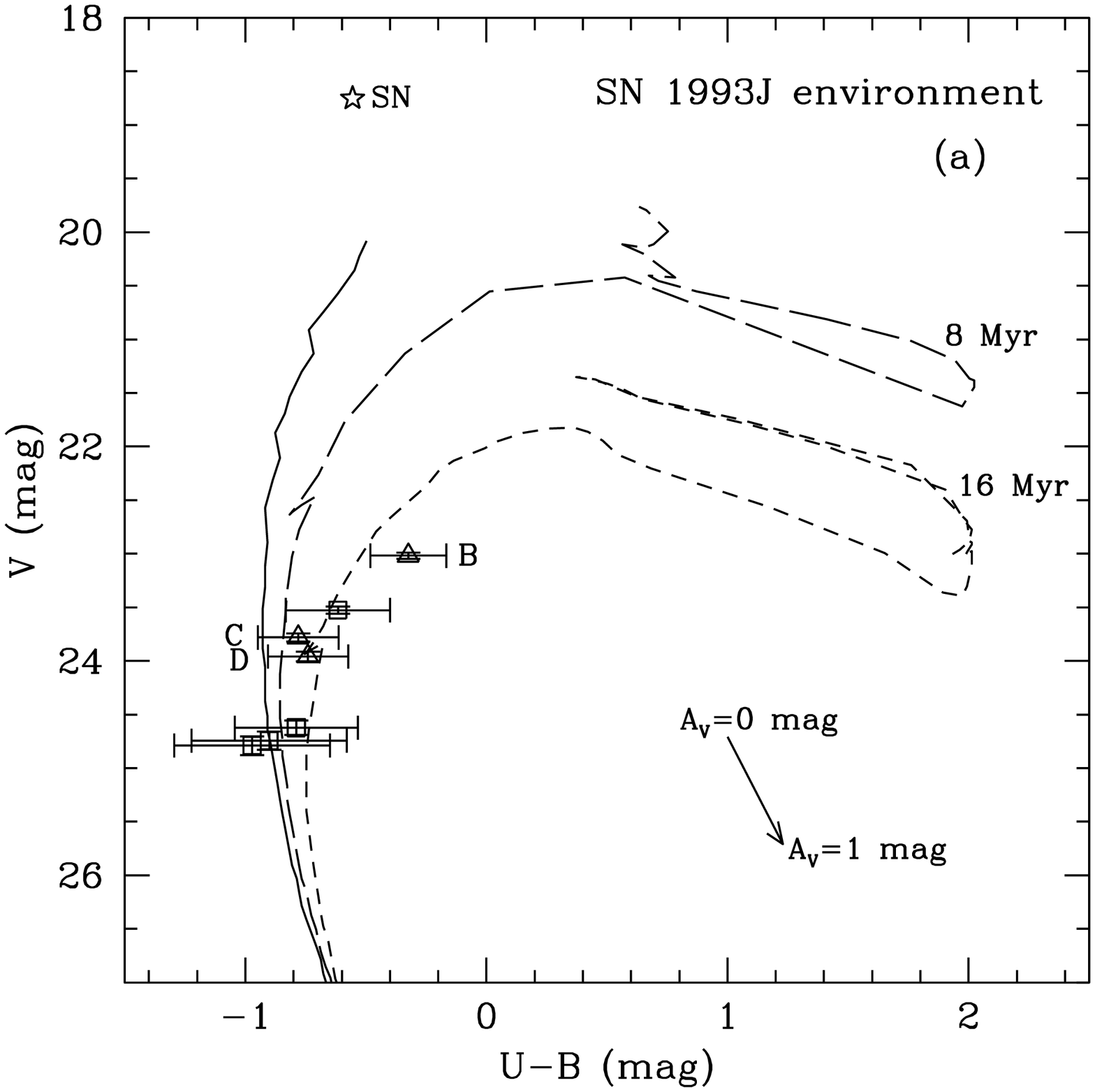}
%\plotfiddle{fg3a.ps}{300pt}{0}{75}{75}{-225}{-130}
\caption{The color-magnitude diagrams for the SN 1993J environment (an area
$\sim 2{\farcs}5$ radius, centered on the SN), measured from the 1995 {\sl
HST\/} images.  {\it (a)} The ($U-B$, $V$) diagram; {\it (b)} the ($B-V$, $V$)
diagram; and {\it (c)} the ($V-I$, $V$) diagram.  In all three diagrams, the
{\it five-pointed star\/} with $V=18.75$ mag is the SN, and Stars A--D are
represented by {\it open triangles\/} and labelled, whereas {\it open
squares\/} represent other stars in the environment.  (In {\it (a)} and {\it
(b)}, these four other blue stars are to the north and northwest of the SN.
In {\it (c)}, the several faint, red stars are distributed throughout the
environment.  The faintest detected 
stars are the reddest, due to the higher sensitivity
of the redder 1995 images.)  The main sequence locus is the {\it solid track};
an 8 Myr isochrone with solar metallicity from Bertelli et al.~(1994), reddened
assuming $A_V = 0.75$ mag and adjusted for the M81 distance from Freedman et
al.~(1994), is shown as the {\it long-dashed track}, and a 16 Myr isochrone is
shown as the {\it short-dashed track}.  A hypothetical progenitor star is
represented by the {\it open circle}.  The reddening vector, following a
Cardelli et al.~(1989) reddening law, is also shown.}
\end{figure}

\clearpage

\begin{figure}
\figurenum{3}
\plotone{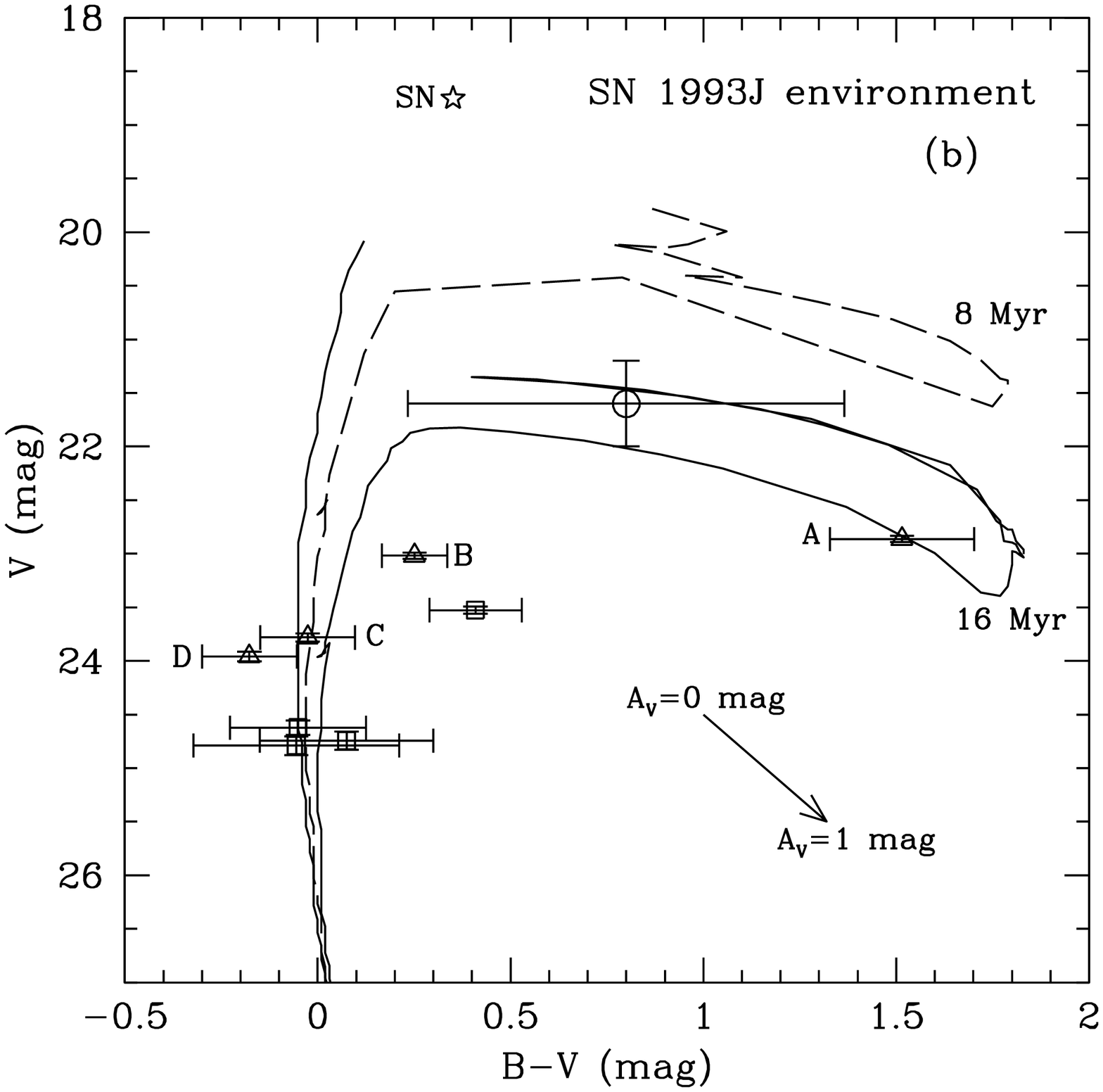}
\caption{(Continued.)}
\end{figure}

\clearpage

\begin{figure}
\figurenum{3}
\plotone{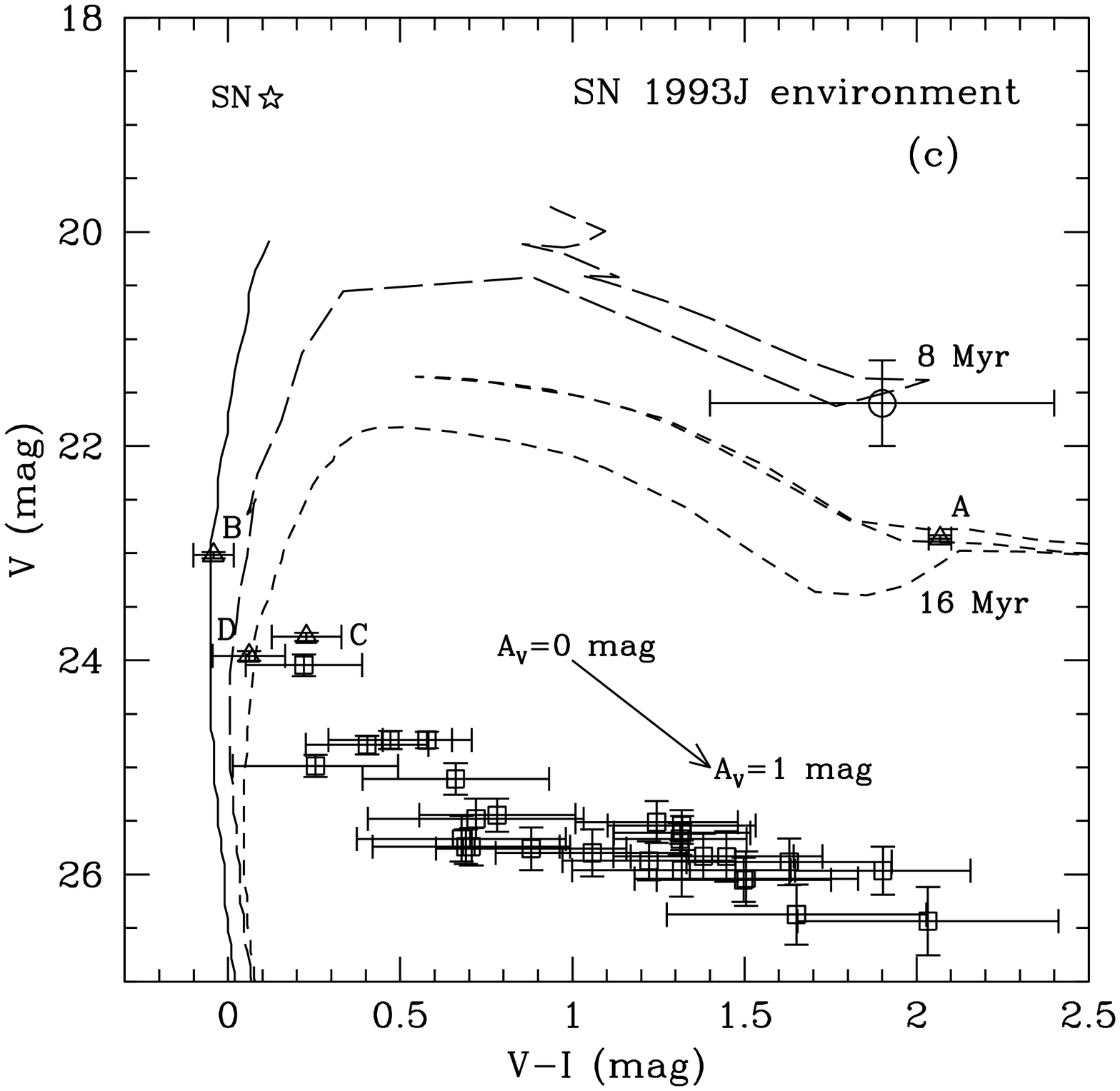}
\caption{(Continued.)}
\end{figure}

\clearpage

\begin{figure}
\figurenum{4}
\plotone{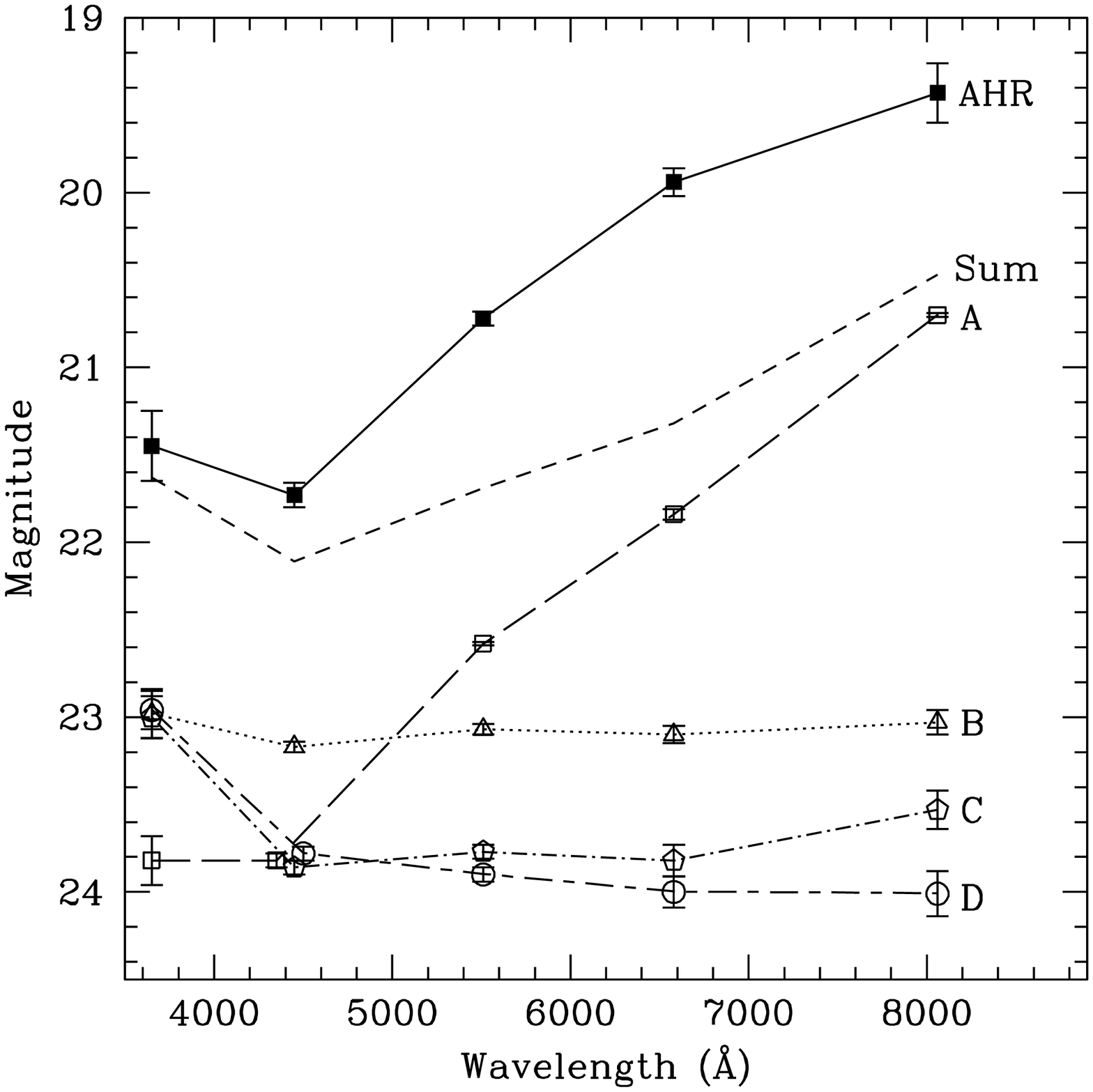}
\caption{The energy distributions of the four bright stars, A--D, within
$2{\farcs}5$ of SN~1993J, compared to the energy distribution for the
SN progenitor from AHR. The {\it short-dashed line\/} shows the sum of all four
stars.}
\end{figure}

\clearpage

\begin{figure}
\figurenum{5}
\plotone{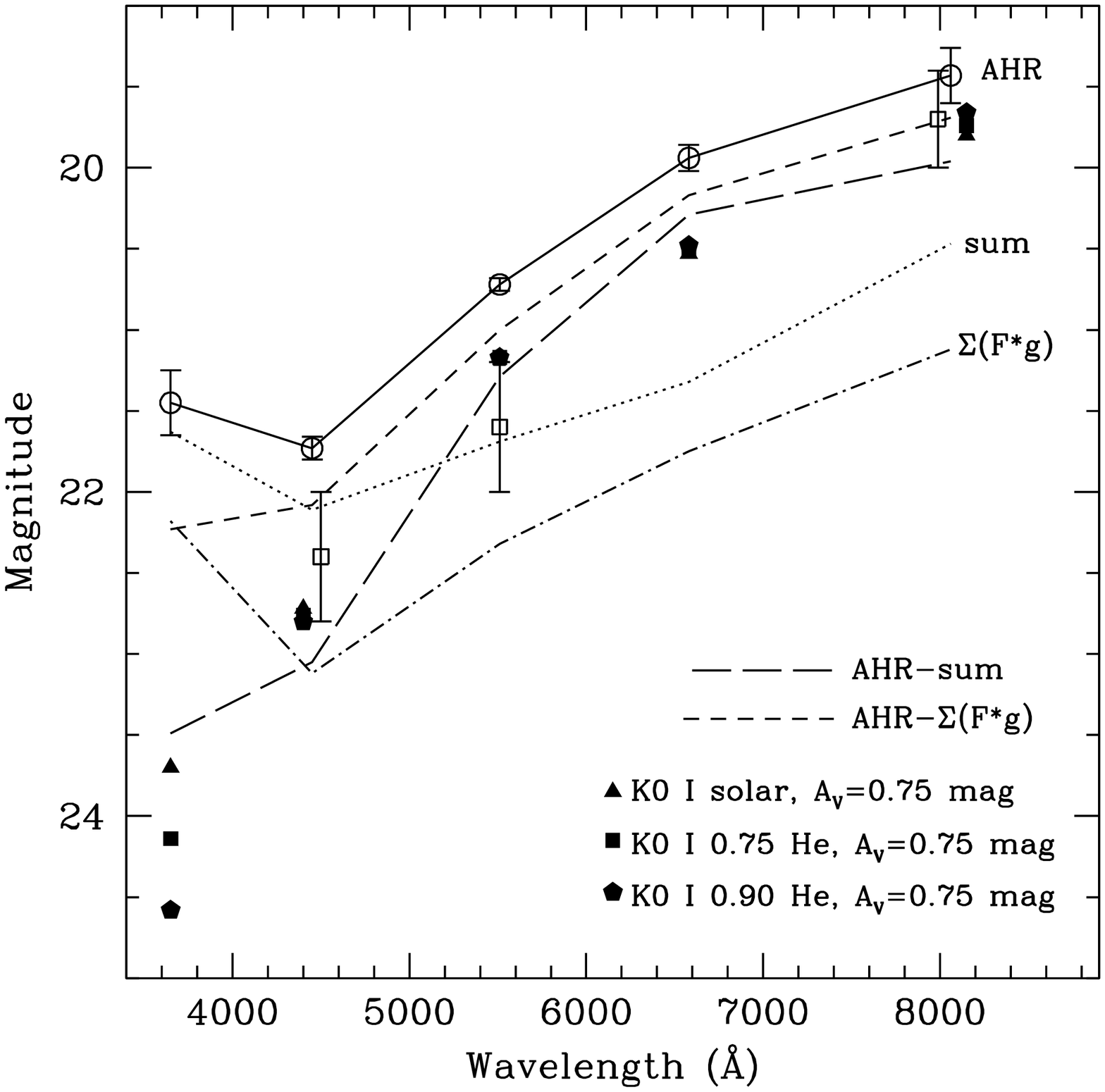}
%\plotfiddle{fg5.ps}{300pt}{0}{75}{75}{-225}{-130}
\caption{The progenitor energy distribution corrected for the contamination
by the four nearby stars. The {\it solid line\/} shows the progenitor
brightness from AHR, using archival ground-based observations; the
{\it dotted line\/} represents the equal-weighted sum of the light from all
four stars; the {\it dot-dashed line\/} represents ${\Sigma}$(F*g), the sum 
of the flux, F, from each of the four stars, weighted by their distance from 
the SN, assuming a one-dimensional Gaussian distribution, g, with 
wavelength-dependent width, $\sigma$ (see text).
The {\it long-dashed line\/} represents the
progenitor energy distribution corrected by the equal-weight sum, while
the {\it short-dashed line\/} represents the distribution corrected by
the Gaussian-weighted sum.  The former is likely a lower limit to the
true progenitor spectrum, while the latter is likely an upper limit.
The {\it open squares\/} represent the light from a fake, hypothetical
progenitor (see text).  The {\it solid symbols\/} show
the energy distribution for model K-type supergiant atmospheres with
a range of possible helium abundances, reddened assuming $A_V = 0.75$ mag.}
\end{figure}

\clearpage

\begin{figure}
\figurenum{6}
%\plotone{fg6a.ps}
\caption{{\it (a)} The F450W ($B$) {\sl HST\/} image from 2001 June
of the SN 1993J environment, after subtraction of the SN via fitting
of a TinyTim PSF (Krist 1995)
within DAOPHOT/ALLSTAR and the addition of a fake progenitor star with
$B \approx 22.4$ mag.
{\it (b)} The F555W ($V$) {\sl HST\/} image from 
2001 June, after SN subtraction and the addition of a fake progenitor 
star with 
$V \approx 21.6$ mag.  
{\it (c)} The F814W ($I$) {\sl HST\/} image from 
2001 June, after SN subtraction and the addition of a fake progenitor 
star with $I \approx 19.7$ mag.  The light echo (Liu et al.~2002; 
Sugerman \& Crotts 2002) is indicated in {\it (a)} and {\it (b)} by an
arrow.}
\end{figure}

%\clearpage

\begin{figure}
\figurenum{7}
%\plotone{fg7a.ps}
\caption{{\it (a)} The F814W ($I$) {\sl HST\/} image from 2001 June
with the fake progenitor star (see Figure 6), after
convolution with a two-dimensional Gaussian with $\sigma=0{\farcs}53$.
{\it (b)} The same as {\it (a)}, except after subtraction of the fake
convolved progenitor via fitting of the TinyTim PSF, also convolved
with the same Gaussian, within ALLSTAR.  The position of Star A is
indicated.}
\end{figure}

\end{document}